\documentclass[10pt,a4paper,twoside]{article}
\usepackage{epsfig}
\usepackage{baltlat6}
\usepackage{array}
\begin{document}
\ \
\vspace{0.5mm}
\setcounter{page}{101}
\vspace{8mm}

\titlehead{Baltic Astronomy, vol.\,??, ???--???, 2011}

\titleb{VAMDC AS A RESOURCE FOR ATOMIC AND MOLECULAR DATA
          AND THE NEW RELEASE OF VALD}

\begin{authorl}
\authorb{F. Kupka}{1} and
\authorb{the VAMDC Collaboration (P.I. M.-L. Dubernet)}{2,3}
\end{authorl}

\hyphenation{Vienna}
\def\bibtex {{\sc Bib}\TeX\ }

\begin{addressl}
\addressb{1}{Faculty of Mathematics, University of Vienna, Nordbergstra{\ss}e 15,  A-1090 \\ Vienna, Austria; 
Friedrich.Kupka@univie.ac.at}
\addressb{2}{LPMAA, UMR7092 CNRS/INP, Universit{\'e} Pierre et Marie Curie, France; 
Marie-Lise.Dubernet-Tuckey@upmc.fr}
\addressb{3}{LUTH, UMR8102 CNRS/INSU, Obs{\'e}rvatoire de Paris, France}
\end{addressl}

\submitb{Received: ???; accepted: ???}

\begin{summary} 
The Virtual Atomic and Molecular Data Centre (VAMDC) (M.L. Dubernet et al.\ 2010, JQSRT 111, 2151) 
is an EU-FP7 e-infrastructure project devoted to building a common electronic infrastructure for the 
exchange and distribution of atomic and molecular data. It involves two dozen teams from six EU member 
states (Austria, France, Germany, Italy, Sweden, United Kingdom) as well as Russia, Serbia, and
Venezuela. Within VAMDC scientists from many different disciplines in atomic and molecular 
physics collaborate with users of their data and also with scientists and engineers from the information 
and communication technology community. In this presentation an overview of the current status of
VAMDC and its capabilities will be provided. In  the second part of the presentation I will focus on one of 
the databases which have become part of the VAMDC platform, the Vienna Atomic Line Data Base (VALD). 
VALD has developed into a well-known resource of atomic data for spectroscopy particularly in astrophysics.
A new release, VALD-3, will provide numerous improvements over its predecessor. This particularly relates
to the data contents where new sets of atomic data for both precision spectroscopy (i.e., with data for 
observed energy levels) as well as opacity calculations (i.e., with data involving predicted energy levels)
have been included. Data for selected diatomic molecules have been added and a new system for data 
distribution and data referencing provides for more convenience in using the upcoming third release of VALD.
\end{summary}

\begin{keywords} atomic data -- molecular data     
\end{keywords}

\resthead{VAMDC as a resource and the new release of VALD}
{F. Kupka and the VAMDC Collaboration (P.I. M.-L. Dubernet)}

\sectionb{1}{INTRODUCTION}

The Virtual Atomic and Molecular Data Centre (VAMDC) has been founded to develop
an interoperable electronic infrastructure for the exchange of atomic and molecular
data. It unites 15 administrative partners who represent 24 teams from six European Union
member states (Austria, France, Germany, Italy, Sweden, United Kingdom) as well as
the Russian Federation, Serbia, and Venezuela. Scientists from a wide variety of disciplines 
in atomic and molecular physics are involved in VAMDC as well as scientists and engineers
from information and communication technology. Since 
many partners of VAMDC have already developed specialized data
bases in their own field and many of them maintain a close connection
to both data producer and data user communities, these resources
were a natural asset of VAMDC from its very beginning.

The key difference of VAMDC to the databases provided by any of its
contributors is that the newly developed electronic infrastructure allows
access to each of these data resources through a single portal.
The effort required by users in finding and retrieving data can thus
be minimized, while the data sets themselves become available to 
a much wider community.

One of the databases integrated into the VAMDC framework is
VALD, the Vienna Atomic Line Data Base (Piskunov et al.\ 1995,
Kupka et al.\ 1999, Ryabchi\-kova et al.\ 1999, Heiter et al.\ 2008).
Developed and maintained at three of the VAMDC partner institutions, 
VALD has served as one of the databases selected for prototype implementations of 
VAMDC. The data collection available to VALD has been greatly enhanced
as a result of very close collaboration with data producers. Through the
development of VAMDC new functionality has been added. This new release,
denoted VALD-3, has now reached a state of development which soon will
permit its official distribution to the user community.

In the following section, the main concepts and capabilities
of VAMDC are described. Its current status and further development
plans are then reviewed. Finally, the most important improvements of
VALD-3 over previous releases are briefly described. A final section 
summarizes these developments and looks forward to new releases
of VAMDC.

\sectionb{2}{CONCEPTS AND CAPABILITIES OF VAMDC}

Atomic and molecular data have been collected and assessed 
in various data\-bases which underpin a wide range of physics in 
applied research and industrial development. Many of them 
have been built to serve specific needs. VALD is one such example
with its own advantages, special tools, and limitations. Heterogeneous
data sets have been collected in different formats and with a varied 
degree of completeness. Specialized extraction tools exist for most 
of the databases, which often significantly contribute to their popularity.
However, this development has also created problems: data often exist
in duplicated form, sometimes also within
a database, which requires non-trivial selection criteria when using them.
Moreover, different user interfaces have to be handled each time a different
database is being accessed. Thus access may be restricted and the
available data may be fragmentary as a result of the collection process. 

The main outcome expected from the VAMDC project is to
\begin{itemize}
\item develop and extend standards for interoperability of resources on atomic and molecular data;
\item implement its concepts for selected databases;
\item allow easy access to a myriad of data resources; 
\item query those resources with dedicated protocols and query languages;
\item allow asynchronous transfer of large amounts of data; 
\item create a safe environment for publishing the latest sets of atomic and molecular data;
\item link data producers more closely with data users.
\end{itemize}
The key benefits expected from using VAMDC are the possibilities to
\begin{itemize}
\item find any type of atomic and molecular data with a single click;
\item have uniform access to the published data;
\item allow cross matching of different data sets; 
\item allow wide access to latest published data.
\end{itemize}

VAMDC clientele include people working in the fields of astrophysics, astronomy,
planetary science, atmospheric science, fusion science, plasma science, the
radiation sciences and in the application of such research in e.g.\ industrial development, for
instance, in the lighting industry. 

The data exchange with databases that are part of VAMDC is organised through
a web-based transport protocol. A user interface and an automatic interface are 
supported. This also allows access of the data through procedures developed
for the international virtual observatory (Quinn et al.\ 2004). Publishing tools of data
producers can access the system in a similar fashion. Mirroring and synchronization of the
capabilities developed within VAMDC ensures a highly reliable service. This is backed
up by an archiving strategy for long-term preservation of the distributed contents.

The transport protocol of VAMDC handles database queries for the status and data content
(registry update), the query and data transfer between the user and the VAMDC portal as 
well as between specific databases and the portal. The transport protocol is self-descriptive 
(XML description of the data sent including units, formats etc.\ based on an extension of
the XSAMS data format of Dubernet et al.\ 2009). It is also efficient for large data sets through 
using compressed binary tables. The interface to each database is fully compatible with the
VAMDC transport on the outside (i.e.\ how the data resource is seen by a VAMDC user), 
but tuned to the specific database on the inside (i.e.\ how VAMDC sees a data resource). 
Functionalities of this interface include converting incoming queries to the internal query
format and converting the database extraction to the transport-compatible VAMDC output
format. Further capabilities are the ability to respond to VAMDC-specific queries such as
registry updates (to keep track of which resources are available at what location) and 
collecting accounting information.

For a more detailed discussion of the scope of the VAMDC project the reader is referred to
Dubernet et al.\ (2010) and to http://www.vamdc.eu/.

\sectionb{3}{STATUS AND DEVELOPMENT OF VAMDC}

An overview on the ``Level 1 Release'' of VAMDC has been provided by Rixon et al.\ (2011).
Achievements during this period included first prototypes of the protocol and XSAMS
data model which were used to develop and provide archive-data services for a small selection
of databases: VALD (Piskunov et al.\ 1995, Kupka et al.\ 1999, Ryabchikova et al.\ 1999, Heiter et al.\ 2008),
XstarDB (Bautista and Kallman 2001), BASECOL (Dubernet et al.\ 2006), and CDMS (M\"uller et al.\ 2005).

Subsequently, protocols and data models have been further refined and a test web-site
has been set up with a registry browser which enables users to access data contents from a much
larger number of databases. In Table~1 we list the databases included in the level 2 release of VAMDC
which are accessible through the registry browser at the time of writing this paper and further databases
for which access is just being prepared. All of them are expected to be available upon completion 
of the level 2 release. Work on further databases is in progress.

\begin{table}[!t]
\begin{center}
\vbox{\footnotesize\tabcolsep=3pt
\parbox[c]{124mm}{\baselineskip=10pt
{\smallbf\ \ Table 1.}{\small\
Databases accessible through the VAMDC Level 2 Release registry browser. Further
information include the VAMDC node hosting the database, and its availability (all databases
including those in preparation (in prep.) will become accessible at the most recent portal level).
References to the original description are listed in column 5. \lstrut}}
\begin{tabular}{ccccl}
\hline
database     &    node                &    status   & portal   &  reference \\
\hline
 BASECOL  & LPMAA                   &  available    & level 2  & Dubernet et al.\ 2006 \\
 CDMS & K\"oln  &  available  & level 2  & M\"uller et al.\ 2005  \\
 CDSD         & IAO (LTS) Russia  &  available  & level 2  & Perevalov \& Tashkun 2008   \\ 
 CHIANTI     & Cambridge / MSSL  &  available    & level 2  & Dere et al.\ 2009     \\
 Ethylen       & Reims               &  available      & level 2  &  see S\&MPO and vadmc.eu      \\
 HITRAN      & UCL                  &  available     & level 2  & Rothman et al.\ 2009      \\
 UDfA (UMIST)   &   Manchester / QUB (Belfast)  &     available  & level 2  & Woodall et al.\ 2007   \\ 
 VALD        &  Uppsala University (mirror)          & available  & level 2  & see text    \\
\hline
  GhoSST   &  Grenoble  & available  & level 1  & Schmitt et al.\ 2009    \\
  Lund data    & Uppsala University & available & level 1  & various  \\
  Methane lines  &  Dijon &  available  & level 1  & see S\&MPO and vadmc.eu  \\
  S\&MPO          & Reims & available  & level 1  & Mikhailenko et al.\    \\
  Spectr-W$^3$     &  RFNC--VNIITF & available  & level 1  & Faenov et al.\ 2002   \\
\hline
  KIDA        &  Bordeaux  &  in prep.  & & kida.obs.u-bordeaux1.fr    \\
  Stark-B    &  Paris-Meudon & in prep.  & & Jevremovi\'c et al.\ 2009  \\
  TipTopBase & IVIC / Cambridge & in prep.  & & Cunto et al.\ 1993   \\
\hline  
\end{tabular}
}
\end{center}
\vskip-4mm  
\end{table}

Detailed references concerning these databases can be found in Dubernet et al.\ (2010)
and also on the VAMDC web site (http://www.vamdc.eu/), the Wiki page of
VAMDC (http://voparis-twiki.obspm.fr/twiki/bin/view/VAMDC/WebHome), and
the VAMDC newsletter available through these resources. After the Level 2 release
a larger user community will be invited to test the VAMDC portal and its capabilities
and the feedback will be used to improve the final release. 

\sectionb{4}{IMPROVEMENTS IN VALD--3}

VALD was created by an international team of researchers (Piskunov et al.\ 1995, 
Kupka et al.\ 1999, Ryabchikova et al.\ 1999, Heiter et al.\ 2008). A lot of its early
development occurred during small workshops at the Institute for Astronomy at the
University of Vienna which became the site of the main server with mirror sites of the
database installed at Uppsala University and the Institute of Astronomy of the Russian
Academy of Sciences in Moscow. Today most of the software development and work 
on integrating molecular data into VALD is performed at Uppsala University. Most of 
the atomic data collection and systematic testing of the data distributed through VALD 
occurs at the Institute of Astronomy of the Russian Academy of Sciences in Moscow.

Ahead of the VALD-3 release, the database contains over 160 line lists and
over 66 million atomic lines provided by all major spectroscopy 
centres across the world. The mirror sites in Vienna, Uppsala, and Moscow 
serve nearly 1500 users from more than 50 countries. On average 30 requests
are being processed per day.

VALD was designed to compile accurate and complete line lists 
for stellar atmospheres and spectroscopy. Line lists included within it are
evaluated to provide a ranking which is used to prefer in the case of duplicated
entries one set of data over another. The database was designed to
be expandable with respect to data and content to allow simple access
through customized extraction software, fast access to individual
data entries, and allow either an overview of parameters from
different sources or, as an alternative choice, extract sets of the
best available data according to data ranking lists and compile
data references for citing original sources.

To this end before adding a data set to the database the data have
to be converted first into a standard format with units which are commonly
used in astrophysics. Multiple extraction layers within VALD allow
accessing and merging of the data and prepare output for different 
applications. Requests can be posed as e-mails or issued through
a web interface. Originally data were sent back only by e-mail, although
as part of VALD-3 a possibility for downloading much larger responses
to user requests will be offered (internally, data output can simply
be streamed into files or standard text input interfaces). 

Several quantities must be known about a line when adding its data to
VALD: central wavelength, species identifier, $\log(gf)$, and energies
as well as total angular momentum quantum numbers of lower and upper
levels of the transition. These are the most crucial data for calculating
absorption lines (in local thermal equilibrium) and they are also used for
deciding on how to merge line lists from different sources. Lines are considered
to be identical, if they originate from the same species, have identical 
total angular momentum quantum numbers for both levels and differ by
less than a threshold with respect to wavelength and energy of both levels.

Further data entries which can be added for a spectral line included within the
VALD database are Land\'e-factors of both energy levels, damping constants
for natural line broadening as well as quadratic Stark effect and Van-der-Waals
broadening, term designations, and information on accuracy or comments on
multiplets, as well as some multipurpose flags.

For spectral lines where both energy levels are experimentally known the
mandatory data entries are usually sufficient to uniquely distinguish them
and avoid erroneous identification of duplicates. Since lines with at least 
one predicted energy level have a very low ranking, such data will be rejected, 
when misidentified with a spectral line with observed energy levels,
which results in the worst case in not putting out a (usually very weak)
predicted line. No false information can be generated this way in an entry
put out as a response to a request to the database. Such a potential loss 
only affects statistical opacity calculations for model atmospheres, since 
for precision spectroscopy data with even just one predicted energy level
is useless because of an unacceptable uncertainty in wavelength of the transition.
It is important to note here that also term designations may be uncertain for
some spectral lines and that their notation can be incoherent among different
sources. Hence, using them as an additional criterion for line identification
was not an option when VALD was originally created. The introduction of
such a consistent notation for the line data stored is still an ongoing project.

The ranking provided by the VALD team uses quality determinations based on error
estimates from original sources, comparisons with existing alternative
sources, and applications in astrophysics (user feedback). A re-ranking
is possible and has occurred on several occasions. General guidelines
for this process are that experimental data are preferred over calculations
(with few exceptions) and data with individual error estimates are considered
more reliable. Line lists from homogeneous sources with high accuracy data are 
given priority over sources which are inhomogeneous or have low quality data.
To add flexibility to this process users can change the ranking for the merging process.

While the first two releases of VALD, dating from 1995 and 1999 (with a subsequent
number of data updates) were accomplished by a rather small team of researchers, 
the VALD-3 release is now being prepared by a much larger group which is a consortium
of scientists from different institutions. Already the core group working in Uppsala, 
Moscow, and Vienna is twice as large and includes U.~Heiter, N.~Piskunov, 
H.C.~Stempels as well as P.~Barklem and O.~Kochukhov at Uppsala University,
R.~Kildiyarova, Yu.~Pakhomov and T. Ryabchikova at Institute of Astronomy of the
Russian Academy of Sciences in Moscow, with both sides supported by further researchers
working primarily on VAMDC, and F.~Kupka, T.~Rank-L\"uftinger, W.W.~Weiss at University
of Vienna, where further researchers have contributed to earlier stages of VALD-3
(L.~Fossati, N.~Nesvacil, M.~Obbrugger, Ch.~St\"utz).
There is also a very close collaboration with many of the leading data producers
in the area of interest for VALD. This originally included the spectroscopy group at the
University of Wisconsin headed by J.E.~Lawler and E.A.~Den~Hartog, the
spectroscopy group at Lund University (H.~Nilsson et al.), the Dream Database team (Bi\'emont et
al.\ 1999), as well as B.~Plez at the Universit\'e de Montpellier and R.L.~Kurucz
at CfA in Harvard, the team of J.S.~Sobeck et al.\ at the University of Texas at Austin,
and the team at Imperial College London, headed by J.C.~Pickering and R.J.~Blackwell-Whitehead.
The list of data provides is still growing at this point. 

In VALD-3, data are still sorted as a function of wavelength and still stored
in a special compressed format allowing semi-direct access to individual
entries. The data stored therein still contains for each spectral line
a species description, wavelength (inside the database consistently in {\AA} for
{\em vacuum conditions}), energies of lower and upper levels ({\em now in cm$^{-1}$}), 
total angular momentum quantum number, oscillator strength in the form of
$\log gf$, Land\'e-factors of both energy levels, damping constants, but
also more information on the {\em accuracy of $\log gf$}, and in addition to the
data reference, a {\em full designation of level and term name}. 

Publishing a new data set in VALD-3 still means adding a new data file to
the existing set, while data description continues to be stored in various
support files (i.e., a list of species, and a configuration file which stores
ranks for every field in each file to allow merging data from different sources).
An important change is that references for each data set are now provided 
in \bibtex format. Indeed, the origin of each entry is now accessible even
for merged line lists such as BELLHEAVY (cf.\ Kurucz 1992). 

In addition to many smaller new lists the New Kurucz Calculations (2006-2010) 
for Fe-peak elements have been included into the data. All these additions allow
much more accurate matching of absorption line spectra of various types of stars.
The model-based selection of lines within VALD, a very popular tool among users, 
has also been improved to make estimates of the contribution to opacities and
predictions of the line strength more accurate and convenient. 




\sectionb{5}{OUTLOOK AND CONCLUSIONS}

In spite of its improvements the upcoming VALD-3 release is subject to a number of 
restrictions: the range of ionization stages is limited (neutral up to 8 times ionized)
and only simple molecules
will be included in VALD-3 (basically diatomics: TiO, CO, CN, CH, C$_2$, FeH).
Generally missing data include collisional transition probabilities and advanced
broadening approximations, among others. Since the VALD consortium has
neither the personnel nor the expertise to fix these deficiencies, even frequent
VALD users may sometimes want to look for alternatives. One possible
solution is to access the VAMDC data resources.

With its very wide range of atomic and molecular databases VAMDC can offer
a common access portal to a large variety of data collections through platform
and database independent methods. It thus can provide benefits for both current
users of the databases participating in VAMDC and to novice users searching
for information on atomic and molecular data who will no longer have to learn
how to use a multitude of interfaces.

A number of conferences will provide opportunities for testing VAMDC and
also offer tutorials on how to use its capabilities. The third annual VAMDC
Conference which will take place in Vienna, Austria, from 21-24 February 2012
(see http://www.vamdc.eu/) is one such opportunity, as is the VAMDC
Regional Workshop and School in Atomic and Molecular Data, which is to 
take place in Belgrade, Serbia, from 7-9 June 2012 
(http://poincare.matf.bg.ac.rs/\~{}andjelka/VAMDC/). It is hoped that in the
future VAMDC will become the reference location in the internet where to
look for atomic and molecular data.

\thanks{ F. Kupka would like to express his gratitude to the LOC
of the 8$^{\rm th}$~SCSLSA for financial support as well as the 
Austrian FWF for funding through project P21742-N16. The help and
suggestions of U.~Heiter and N.~Mason who carefully proofread 
this manuscript are greatly appreciated. This work
has been presented as part of the VAMDC collaboration. VAMDC
is funded under the ``Combination of Collaborative Projects and 
Coordination and Support Actions'' Funding Scheme of The Seventh 
Framework Program. Call topic: INFRA-2008-1.2.2 Scientific Data 
Infrastructure. Grant Agreement number: 239108.}

\References

\refb Bautista, M.A., Kallman, T.R. 2001, Astrophys. J. Suppl. 134, 139

\refb Bi\'emont, E., Palmeri, P., Quinet, P. 1999, Astrophys. J. Suppl. 635, 2691

\refb Cunto, W., Mendoza, C., Ochsenbein, F., Zeippen, C. 
1993, A\&A 275, L5 

\refb Dere, K.P., Landi, E., Young, P.R. et al.\ 2009, A\&A 498, 915 

\refb Dubernet, M.L., Grosjean, A., Flower, D. et al.\ 2006, 
Ro-vibrational Collisional Excitation Database BASECOL‚ $\langle$http://basecol.obspm.fr/$\rangle$, 
in Proceedings of the Joint Meeting ITC14 and ICAMDATA 2004, Toki, Japan, J. Plasma Fusion Res. Ser. 7, 356

\refb Dubernet, M.L., Humbert, D., Clark, R.E.H. et al.\ 
2009, XSAMS: XML schema for Atomic, Molecular and Solid Data. In: Dubernet, M.L., Humbert, D.,
Ralchenko, Yu., editors. $\langle$http://www-amdis.iaea.org/xml/$\rangle$, Version 0.1,
Sep\-tember 2009.

\refb Dubernet, M.L., Boudon, V., Culhane, J.L. et al.\ 2010, 
Jour. Quant. Spectr. Rad. Transfer, 111, 2151

\refb Faenov, A.Y., Magunov, A.I., Pikuz, T.A. et al.\ 2002,
Spectr-W-3 online database on atomic properties
of atoms and ions. AIP Conf. Proc. 636, 253 

\refb Heiter, U., Barklem, P., Fossati, L. et al.\ 2008,
VALD -- an atomic and molecular database for astrophysics,
Journal of Physics Conference Series 130, 1

\refb Jevremovi\'c, D., Dimitrijevi\'c, M.S., Popovi\'c, L.\u{C} et al.\ 2009,
New Astron. Rev. 53, 222 

\refb Kupka, F., Piskunov, N., Ryabchikova, T.A. et al.\ 1999, 
A\&A Suppl., 138, 119

\refb Kurucz, R.L. 1992, Rev. Mex. Astron. Astrof. 23, 45

\refb Mikhailenko, S., Barbe, A., Babikov, Y., Tyuterev, V.G, 
S\&MPO— a databank and information system for ozone spectroscopy on 
the WEB, $\langle$http://smpo.iao.ru/$\rangle$. 

\refb M\"uller, H.S.P., Schl\"oder, F., Stutzki, J., Winnewisser, G. 2005, J. Mol. Struct. 742, 215

\refb Perevalov, V.I., Tashkun, S.A. 2008, CDSD-296 (Carbon Dioxide Spectroscopic 
Databank): updated and enlarged version for atmospheric applications. 
In: 10th HITRAN database conference, Cambridge, MA, USA; 
fourth assessment report of the intergovernmental panel 
on climate change. Cambridge, UK: Cambridge University Press.

\refb Piskunov, N.E., Kupka, F., Ryabchikova, T.A. et al.\ 1995, 
A\&A Suppl., 112, 525

\refb Quinn, P., Barnes, D., Csabai, I. et al.\ 2004, The International Virtual Observatory Alliance: 
recent technical developments and the road ahead, SPIE 5493, 137

\refb Rixon, G., Dubernet, M.L., Piskunov, N. et al.\ 2011, 
in 7th International Conference on Atomic and Molecular Data and Their Applications 
-- ICAMDATA--2010, AIP Conf.\ Proc.\ 1344, 107

\refb Rothman, L.S., Gordon, I.E., Barbe, A. et al.\ 2009, JQSRT 110, 533

\refb Ryabchikova, T.A., Piskunov, N.E., Stempels, H.C., et al.\ 1999, 
in Proc.\ of the 6th Int.\ Coll.\ on Atomic Spectra and Oscillator Strengths, Victoria BC, Canada,
Phys. Scripta, T83, 162

\refb Schmitt, B.P., Volcke, E., Quirico, O. et al.\ 2009,
GhoSST: the Grenoble astrophysics and planetology solid 
spectroscopy and thermodynamics database service: ‘‘RELEVANT 
Database’’, see $\langle$http://ghosst.obs.ujf-grenoble.fr/$\rangle$. 

\refb Woodall, J., Ag\'undez, M., Markwick-Kemper, A.J., Millar, T.J. 2007, A\&A 466, 1197 

\end{document}